\documentclass{amsart}
\usepackage{amssymb}

\begin{document}

\title[Riemannian-Lorentzian metrics]{Variational equations on mixed Riemannian-Lorentzian metrics}
\author{Thomas H. Otway}
\address{Department of Mathematics, Yeshiva University, 500 W 185th Street, New York, New York 10033, U.S.A}
\email{otway@yu.edu}
\date{}
\begin{abstract}
A class of elliptic-hyperbolic equations is placed in the context
of a geometric variational theory, in which the change of type is
viewed as a change in the character of an underlying metric. A
fundamental example of a metric which changes in this way is the
extended projective disc, which is Riemannian at ordinary points,
Lorentzian at ideal points, and singular on the absolute. Harmonic
fields on such a metric can be interpreted as the hodograph image
of extremal surfaces in Minkowski 3-space. This suggests an
approach to generalized Plateau problems in 3-dimensional
space-time via Hodge theory on the extended projective disc.
Analogous variational problems arise on Riemannian-Lorentzian flow
metrics in fiber bundles (twisted nonlinear Hodge equations), and
on certain Riemannian-Lorentzian manifolds which occur in
relativity and quantum cosmology. The examples surveyed come with
natural gauge theories and Hodge dualities. This paper is mainly a
review, but some technical extensions are proven.
\textit{MSC2000}: 35M10, 53A10, 83C80
\medskip

\noindent\emph{Key words}: signature change, projective disc,
Minkowski 3-space, equations of mixed type, nonlinear Hodge
equations
\end{abstract}
\maketitle

\section{Introduction: the projective disc}

\noindent\emph{In a small circle of paper, you shall see as it
were an epitome of the whole world.}

\medskip

\qquad \qquad Giambattista Della Porta, 1589, on the \emph{camera
obscura}

\bigskip

Analysis on Beltrami's projective disc model for hyperbolic space
is in one sense very old mathematics. Beltrami introduced the
projective disc in 1868 as one of the earliest Euclidean models
for non-Euclidean space \cite {B1}; see also \cite{B2}, \cite{S2}.
But it also arises in the context of some new mathematics related
to variational problems in Minkowski space and Hodge theory on
pseudo-Riemannian manifolds. In practice Beltrami's construction
amounts to equipping the unit disc centered at the origin of
coordinates in $\mathbb{R}^2$ with the distance function
\begin{equation}\label{Belt}
    ds^{2}=\frac{\left( 1-y^{2}\right) dx^{2}+2xydxdy+\left(
1-x^{2}\right) dy^{2}}{\left( 1-x^{2}-y^{2}\right) ^{2}}.
\end{equation}
Integrating $ds$ along geodesic lines in polar coordinates, we
find that the distance from any point in the interior of the unit
disc to the boundary of the disc is infinite, so the unit circle
becomes the \emph{absolute}: the curve at projective infinity;
\emph{c.f.} \cite{He}, Sec.\ 9.1.

It is natural to ask how to interpret the so-called \emph{ideal}
points in the complement of the unit disc in $\mathbb{R}^2.$ It
has been known for a long time that such points are not merely
allowed by the projective disc model in a formal sense, but are
actually useful in classical geometric constructions. For example,
tangent lines to the unit disc can be used to characterize
orthogonal lines within the disc, and certain families of
translated lines inside the disc attain their simplest
representation as a rotation about an ideal point. Although these
classical geometric operations on ideal points are well known
\cite{S1}, \cite{Le}, geometric analysis on domains which include
ideal points is still very incompletely understood.

In this review we use Beltrami's model as a point of reference
from which to survey aspects of geometric variational theory on
\emph{mixed Riemannian-Lorentzian} domains, on which the signature
of the metric changes sign along a smooth hypersurface. (The
metric underlying (\ref{Belt}) changes from Riemannian to
Lorentzian on the Euclidean unit circle.) In Sec.\ 2 we consider
variational problems which reduce to the existence of harmonic
fields on the extended projected disc $-$ that is, solutions to
the \emph{Hodge equations}
\begin{equation}\label{hodge}
    d\alpha=\delta \alpha = 0
\end{equation}
on a domain of $\mathbb{R}^2$ which includes the closed unit disc
as a proper subset and is equipped with the Beltrami metric
(\ref{Belt}). Here $d$ is the exterior derivative, with formal
adjoint $\delta,$ acting on a 1-form $\alpha.$ The crucial
technical problem for the equations of Sec.\ 2 is the existence of
solutions to boundary-value problems; this question is addressed
in Sec.\ 3. Sections 4 and 5 are concerned with the related topics
of duality and gauge invariance. These topics motivate the study
of gauge-invariant, potentially elliptic-hyperbolic systems in
fiber bundles. In Sec.\ 6 we briefly review similar systems that
arise in relativity and cosmology, and the relation of those
systems to other elliptic-hyperbolic equations on manifolds which
have been studied by mathematicians.

On the extended projective disc the Hodge equations are no longer
uniformly elliptic, but are elliptic on ordinary points inside the
unit disc, hyperbolic on ideal points in the
$\mathbb{R}^2$-complement of the disc, and parabolic on the unit
circle, which is a singularity of the manifold. One might wonder
why anyone would want to study such a peculiar system. There are
at least two motivations for doing so:

\emph{i)} to learn what the geometry of the Beltrami disc reveals
about elliptic-hyperbolic partial differential equations, and

\emph{ii)} to learn what elliptic-hyperbolic partial differential
equations on the Beltrami disc reveal about the geometry of
space-time.

Regarding the first motivation, there is no canonical way to
decide what constitutes a natural boundary-value problem for an
equation that changes from elliptic to hyperbolic type on a smooth
curve. Historically, physical analogies have been the main tool,
chiefly analogies to the physics of compressible flow \cite{Be}.
However, it is also possible to approach the problem using a
geometric analogy, which we will briefly describe.

The highest-order terms of any linear second-order partial
differential equation on a domain $\Omega \subset \mathbb{R}^2$
can be written in the form
\begin{equation}\label{secord}
    Lu=\alpha \left( x,y\right) u_{xx}+2\beta (x,y)u_{xy}+\gamma
    (x,y)u_{yy},
\end{equation}
where $(x,y)$ are coordinates on $\Omega;$ $\alpha,$ $\beta,$ and
$\gamma$ are given functions; $u=u\left(x,y\right).$
Traditionally, the question of whether the equation is elliptic,
hyperbolic, or parabolic has been identified with the question of
whether the discriminant
\begin{equation}\label{discr}
    \Delta \left( x,y\right) =\alpha \gamma -\beta ^{2}
\end{equation}
is respectively positive, negative, or zero. If the discriminant
is positive on part of $\Omega $ and negative elsewhere on $\Omega
,$ then the equation associated with the operator $L$ is said to
be of \textit{mixed elliptic-hyperbolic type}. The curve on which
the equation changes type is called the \emph{parabolic line} or,
borrowing the terminology of fluid dynamics, the \emph{sonic
curve}. A simple example of an elliptic-hyperbolic equation is the
\emph{Lavrent'ev-Bitsadze equation}
\[
sgn(y)u_{xx}+u_{yy}=0,
\]
for which the parabolic line is the $x$-axis.

An alternative approach is to replace the class of differential
operators $L$ on a single domain of $\mathbb{R}^2$ with the single
\textit{Laplace-Beltrami operator,}
\[
\mathcal{L}_g u=\frac{1}{\sqrt{\left| g\right| }}\frac{\partial
}{\partial x^{i}}\left( g^{ij}\sqrt{\left| g\right|
}\frac{\partial u}{\partial x^{j}}\right) ,
\]
on a class of domains for which $g_{ij}$ represents the local
metric tensor.

For example, the Lavrent'ev-Bitsadze equation can be associated to
the Laplace-Beltrami operator on a metric which is Euclidean above
the $x$-axis and Minkowskian below the $x$-axis. Elliptic
operators will be associated to Riemannian metrics; wave operators
will be associated to Lorentzian metrics. In this classification,
the type of a linear second-order equation is not a function of
the associated linear operator at all; that operator is always the
Laplace-Beltrami operator. Rather, the type of the equation is a
feature of the metric tensor on the underlying surface. Any change
in the signature which results in a change in sign of the
determinant $g$ will change the Laplace-Beltrami operator on the
metric from elliptic to hyperbolic type. The Laplace-Beltrami
operator on surface metrics for which such a change occurs along a
smooth curve will correspond to planar elliptic-hyperbolic
operators in local coordinates. However, any curve on which the
change of type occurs will necessarily represent a singularity of
the metric tensor, as the determinant $g$ will vanish along that
curve.

According to this point of view, in order to decide which
boundary-value problems are natural for a second-order linear
elliptic-hyperbolic equation on $\mathbb{R}^2,$ one should study
the geometry of the underlying pseudo-Riemannian metric. For
example, characteristic lines for the Hodge equations on extended
$\mathbb{P}^2$ can be interpreted as polar lines for a chord of
the projective disc, and this suggests a natural class of
boundary-value problems \cite{O10}.

However, in this review we focus on the second motivation for
studying harmonic fields on the extended projective disc. This is
the opportunity that they afford to revisit, from a different
point of view, work done by Chao-Hao Gu in the 1980's on the
existence of extremal surfaces in Minkowski 3-space
$\mathbb{M}^{2,1}$ \cite{Gu1}-\cite{Gu5}, and similar variational
problems which arise in the context of fiber bundles.

\section{The geometric variational problem}

The area functional for a smooth surface $\Sigma$ in
$\mathbb{M}^{2,1}$ having graph $z=f\left(x,y\right)$ is given by

\[
A=\int\int_\Sigma\sqrt{\left|1-f_x^2-f_y^2\right|}\,dxdy.
\]
The surface $\Sigma$ is time-like when $f_x^2+f_y^2$ exceeds unity
and space-like when $f_x^2+f_y^2$ is exceeded by unity.
Introducing Lagrange's notation $p=f_x,\,q=f_y,$ the boundary
between the space-like and time-like surfaces is the unit circle
centered at the origin of coordinates in the $pq$-plane.

A necessary condition for $\Sigma$ to be extremal on
$\mathbb{M}^{2,1}$ is that its graph $f\left(x,y\right)$ satisfy
the minimal surface equation in the form \cite{Gu2}
\begin{equation}\label{quasi}
    \left(1-p^2\right)q_y+2pqp_y+\left(1-q^2\right)p_x=0.
\end{equation}
This is a quasilinear partial differential equation which is
elliptic for space-like surfaces and hyperbolic for time-like
surfaces. We can linearize this equation by the method of Legendre
(\cite{CH}, Sec.\ I.6.1), applying the transformation
$z=px+qy-\varphi\left(p,q\right),$ $x=\varphi_p,\,y=\varphi_q.$ We
obtain the linear equation
\[
\left(1-p^2\right)\varphi_{pp}-2pq\varphi_{pq}+\left(1-q^2\right)\varphi_{qq}=0.
\]
Adopting homogeneous coordinates $\left(u,v,w\right)$ for $w\ne
0,$ we eventually obtain the equation \cite{Gu3}
\begin{equation}\label{exp2}
    \left[\left(1-p^2\right)\psi_p\right]_p-2pq\psi_{pq}+\left[\left(1-q^2\right)\psi_q\right]_q=0,
\end{equation}
where $p=-u/w$ and $q=-v/w.$ Equation (\ref{exp2}) can be
interpreted as the Laplace-Beltrami equation on the extended
projective disc $\mathbb{P}^2;$ \emph{c.f.} \cite{O10} and
references therein.

In models for more general classes of surfaces (\emph{e.g.},
Examples 2-4 from Sec.\ 2 of \cite{SS2}), the gradient of $f$ may
be replaced by the $d$-closed 1-form $\omega =
\omega_1dx+\omega_2dy.$ In particular, consider a space-like
surface $\Sigma$ embedded in $\mathbb{M}^{2,1}.$ The variational
equations of the area functional can be written in the form
\begin{equation}\label{H1}
    \left[\left(1-\omega_2^2\right)\omega_1\right]_x+\omega_1\omega_2\left(\omega_{1y}+\omega_{2x}\right)+\left[\left(1-\omega_1^2\right)\omega_2\right]_y=0,
\end{equation}

\begin{equation}\label{H2}
    \omega_{2x}-\omega_{1y}=0.
\end{equation}
These are the equations for an extremal surface which is
space-like in some regions of $\mathbb{M}^{2,1}$ and time-like in
others. However, the surface that it describes is singular on the
circle $\omega_1^2+\omega_2^2=1.$

If the Gaussian curvature of the surface is nonvanishing, then
\[
\omega_{1x}\omega_{2y}-(\omega_{1y})^2=\omega_{1x}\omega_{2y}-\omega_{1y}\omega_{2x}\ne
0
\]
and the nonlinear system can be linearized by a Legendre
transformation to obtain
\begin{equation}\label{P1}
    \left[\left(1-x^2\right)u_1\right]_x-\left(xyu_1\right)_y-\left(xyu_2\right)_x+\left[\left(1-y^2\right)u_2\right]_y=0,
\end{equation}

\begin{equation}\label{P2}
    u_{1y}-u_{2x}=0.
\end{equation}
This system has a geometric interpretation as the Hodge equations
on the extended projective disc.

From this point of view, solving the Hodge equations locally on
extended $\mathbb{P}^2$ is a way to approach generalized Plateau
problems in Minkowski 3-space. For example, suppose one were to
study the existence question for a class of area-extremizing
surfaces in $\mathbb{M}^{2,1}.$ The first step in such a program
might be an existence theorem for a class of \emph{Dirichlet
problems} for eqs.\ (\ref{P1}), (\ref{P2}): The hypothesis of a
sufficiently smooth vector-valued function $h$ prescribed on all
or part of a boundary curve in extended $\mathbb{P}^2$ would be
shown to imply the existence of a 1-form satisfying the Hodge
equations in the interior and equalling $h$ on the boundary.

Unfortunately, a good Hodge theory on extended $\mathbb{P}^2$
would not in itself provide a good variational theory even for
stationary hypersurfaces in $\mathbb{M}^{2,1}.$ One reason is that
the Legendre transformation may itself introduce singularities. It
is known that, in the elliptic region, such singularities can only
occur on at most a point set if the image is the system
(\ref{P1}), (\ref{P2}) \cite{O8}. However, higher-order
singularities in the parabolic and hyperbolic regions of the
equations are possible and, based on analogies to the equations of
gas dynamics \cite{CF}, expected. Moreover, although the Legendre
transformation makes the equation simpler, it makes boundary
conditions more complicated, so the interpretation of
boundary-value problems for harmonic 1-forms on extended
$\mathbb{P}^2$ in terms of extremal hypersurfaces in
$\mathbb{M}^{2,1}$ is not generally straightforward.

Nevertheless, the Dirichlet problem for a quasilinear
elliptic-hyperbolic system, having a line singularity on the
parabolic curve, is sufficiently formidable that the approach via
linearization remains the one with the most apparent promise. (For
a possible alternative, see \cite{Ts} and references therein; for
an alternative in the particular case studied by Gu, see
\cite{Li}.)

\section{Boundary-value problems}

Both of the two motivations for studying the Hodge equations on
extended $\mathbb{P}^2$ begin with the existence question for
boundary-value problems: under what conditions on the boundary can
we expect solutions in the interior? The obstruction to answering
this question is the problematic nature of boundary-value problems
for elliptic-hyperbolic equations.

The systematic study of partial differential equations which
change from elliptic to hyperbolic type on a smooth hypersurface
began in 1923 with the famous memoir of Tricomi \cite{Tr}
concerning the equation
\begin{equation}\label{tricomi}
    yu_{xx} + u_{yy}=0.
\end{equation}
Tricomi's work was extended in the early 1930s by
Cinquini-Cibrario \cite{Ci} to include the equation

\begin{equation}\label{cibrario}
    xu_{xx} + u_{yy}=0.
\end{equation}
The study of these equations led to the establishment of two
classes of normal forms, to which any linear elliptic-hyperbolic
equation of second order can be reduced near a point (\emph{c.f.}
\cite{Bi}), and some appropriate boundary-value problems for such
forms. In particular, the \emph{Tricomi problem} for
(\ref{tricomi}) or (\ref{cibrario}) prescribes the solution on the
part of the boundary consisting of characteristic curves; see,
\emph{e.g.}, Sec.\ 12.1 of \cite{G} for a discussion. Tricomi
problems for the scalar case of the system (\ref{P1}), (\ref{P2})
were solved by Hua and his students in the 1980s \cite{Hu},
\cite{JC}.

The boundary-value problems formulated by Tricomi and
Cinquini-Cibrario were \emph{open} in the sense that the boundary
conditions are prescribed on a proper subset of the boundary.
\emph{Closed} problems, in which data are prescribed on the entire
boundary, have only recently been studied in a systematic way
\cite{LMP} see also \cite{Pa4}. Closed problems were formulated as
early as 1929 by Bateman \cite{Ba}, but in the informal style
typical of British applied mathematics of that period. An isolated
result, the existence of weak solutions to the closed Dirichlet
problem for the scalar Tricomi equation, was proven by Morawetz in
1967 \cite{M3}; see also \cite{Pi}. Some special cases are treated
in \cite{Gu1}, \cite{MaTa}, \cite{O12}, and \cite{T}.

The closed boundary-value problems in \cite{LMP} are formulated
for a class of equations having the form
\begin{equation}\label{t-type}
    K(y)u_{xx}+u_{yy}=0,
\end{equation}
where the \emph{type-change function} $K(y)$ is a continuously
differentiable function satisfying certain technical properties,
the most important of which are $K(0)=0$ and $yK(y) > 0$ for $y
\neq 0.$ In the simplest special case, $K(y)=y,$ (\ref{t-type})
reduces to the Tricomi equation (\ref{tricomi}). For this reason,
equations of the form (\ref{t-type}) are said to be of
\emph{Tricomi type.} Similarly, eq.\ (\ref{cibrario}) is the
simplest example of an equation of \emph{Keldysh type} $-$ that
is, an equation of the form \cite{Ke}
\begin{equation}\label{keldysh}
    K(x)u_{xx}+u_{yy}=0,
\end{equation}
where $K(x)$ is generally taken to be a continuously
differentiable function such that $K(0)=0$ and $xK(x) > 0$ for $x
\neq 0.$ Equations of this kind, with various lower-order terms,
have arisen in the study of transonic gas dynamics and, recently,
in optics; see, \emph{e.g.,} Sec.\ 3 of \cite{CK} and Sec.\ 4 of
\cite{MaTa}, respectively. If the components of the vector
$\left(u_1,u_2\right)$ are sufficiently differentiable, then the
system (\ref{P1}), (\ref{P2}) in polar coordinates can be
represented as a single equation of Keldysh type. Sufficient
conditions for the existence of unique solutions to closed
boundary-value problems for equations having the general form
(\ref{keldysh}) are virtually unknown.

The known results on the existence of solutions to boundary-value
problems for (\ref{P1}), (\ref{P2}) can be summarized as follows:

Let $\Omega$ be the domain formed by the polar lines of a chord of
the unit disc in extended $\mathbb{P}^2.$ Then there exists a weak
solution on $\Omega$ with data prescribed on the
non-characteristic part of the boundary. In fact, we can deform
the chord in such a way that there is also a non-characteristic,
explicitly hyperbolic boundary on which data are prescribed,
providing a mild monotonicity condition is met. Solutions lie in a
weighted function space. The weak solution is strong (in the sense
of Sec.\ 3.2, below) if we round off the sharp points on the
boundary and perturb the lower-order terms. These assertions are
stated precisely and proven in Sec.\ II of \cite{O10}. Although
they are formulated in \cite{O10} in the context of projective
geometry, a (different) boundary-value problem for an analogous
domain in the scalar case has been formulated in the context of
Minkowski geometry; compare Fig.\ 1 of \cite{Gu4} with Sec.\ 4,
Fig.\ 2 of \cite{Hu}.

However, we show in the following section that the closed
Dirichlet problem for twice-continuously differentiable solutions
of (\ref{P1}), (\ref{P2}) is over-determined on the hyperbolic
boundary. Note the odd terminology: Hyperbolic \emph{points} are
ordinary points lying inside the unit disc, and thus lie in the
elliptic region of the eqs.\ (\ref{P1}), (\ref{P2}). The
hyperbolic \emph{boundary} is the portion of the domain boundary
on which the discriminant (\ref{discr}) of the equation is
negative; so the hyperbolic boundary for (\ref{P1}), (\ref{P2})
lies outside the unit disc and consists of ideal points rather
than hyperbolic points.

\subsection{The nonexistence of classical solutions}

Expressed in polar coordinates, eq.\ (\ref{exp2}) assumes the form
\begin{equation}\label{pol0}
    \left(1-r^2\right)\phi_{rr}+\frac{1}{r^2}\phi_{\theta\theta}+\left(\frac{1}{r}-2r\right)\phi_r=0,
\end{equation}
for $\phi = \phi\left(r,\theta\right),$ provided $r\ne 0.$ We
define eq.\ (\ref{pol0}) over a sector
\[
\Omega_s=\lbrace\left(r,\theta\right)|0<r_0\leq r\leq r_1,\,
\theta_1\leq\theta\leq\theta_2\rbrace,
\]
where $\theta_2-\theta_1<\pi$ and the interior of $\Omega_s$
includes a segment of the line $r=1.$ For convenience we will take
$\theta_1$ to be negative and $\theta_2$ to be positive.

Writing eq.\ (\ref{pol0}) in the equivalent form
\begin{equation}\label{pol3}
    r^2\left(1-r^2\right)\phi_{rr}+\phi_{\theta\theta}+r\left(1-2r^2\right)\phi_r=0,
\end{equation}
we show that the Dirichlet problem for this equation on a typical
domain has a unique solution if data are given on only the
non-characteristic portion of the boundary. This will imply that
the Dirichlet problem is over-determined if data are prescribed on
the entire boundary.

Define the set $\Omega^+\subset\Omega_s,$ where
\[
\Omega^+=\left\lbrace\left(r,\theta\right)\in\mathbb{R}^2|r_0<\varepsilon\leq
r<1,-\theta_0\leq \theta \leq \theta_0\right\rbrace.
\]
We will choose the domain $\Omega$ of eq.\ (\ref{pol3}) to be the
region enclosed by the annular sector $\Omega^+$ and the
intersecting lines tangent to the points
$\left(r,\pm\theta\right)=\left(1,\pm\theta_0\right).$

Precisely, let
$\left(r,\pm\theta\right)=\left(1,\pm\theta_0\right)$ be the two
points of intersection of the line $x=x_0,$ $x_0\in
\left(\varepsilon,1\right),$ with the unit circle centered at the
origin of coordinates in $\mathbb{R}^2.$ Let $\Omega_0$ be the
triangular region bounded by the vertical chord $\gamma_0$ given
in Cartesian coordinates by
\[
\gamma_0=\left\lbrace
\left(x,y\right)\in\mathbb{R}^2|x=x_0,y^2\leq 1-x_0^2\right\rbrace
\]
and the two polar lines of $\gamma_0.$ (Recall that the
\emph{polar} lines of a chord are the tangent lines to the unit
circle at its two points of intersection with the chord.) Then
$\Omega=\Omega^+\cup\Omega_0.$

In the following we denote by $\Omega^-$ the subdomain of
$\Omega_0$ consisting of ideal points, and by $\nu$ the arc of the
unit circle lying between the points
$\left(r,\pm\theta\right)=\left(1,\pm\theta_0\right).$ Because
$\varepsilon$ is a fixed number greater zero, the mapping from the
region $\Omega$ in the polar $r\theta$-plane (the Cartesian
$xy$-plane) to its image in the Cartesian $r\theta$-plane is
well-defined, and we shall call the image $\Omega$ as well.

\bigskip

\textbf{Theorem 1}. \emph{Considering (\ref{pol3}) as an equation
on the subdomain $\Omega$ of the Cartesian $r\theta$-plane, any
solution $\phi\in C^2\left(\Omega\right)$ of (\ref{pol3}) taking
values $f\left(r,\theta\right)\in C^2\left(\partial\Omega\right)$
on the boundary segment $\partial\Omega\backslash\Omega^-$ is
unique.}

\bigskip

\emph{Proof}. The method of proof has been applied to other
elliptic-hyperbolic equations \cite{Mw}, \cite{M1}, \cite{MSW}.
Suppose that there are two solutions satisfying the boundary
conditions. Subtraction yields a solution, which we also denote by
$\phi,$ satisfying homogeneous boundary conditions. Define the
functional
\[
I=\int^{\left(r,\theta\right)}\psi_1d\theta+\psi_2dr,
\]
where
\begin{equation}\label{psi1}
    \psi_1=r^2\left(1-r^2\right)\phi_r^2-\phi_\theta^2
\end{equation}
and
\begin{equation}\label{psi2}
    \psi_2=-2\phi_r\phi_\theta,
\end{equation}
on the Cartesian $r\theta$-plane. Because
\[
\psi_{2\theta}-\psi_{1r}=
\]

\begin{equation}\label{multipl}
    -2\phi_r\left[r^2\left(1-r^2\right)\phi_{rr}+\phi_{\theta\theta}+r\left(1-2r^2\right)\phi_r\right]=0,
\end{equation}
there is a function $\chi\left(r,\theta\right)$ such that
$\chi_\theta=\psi_1$ and $\chi_r=\psi_2.$

The first step is to show that $\phi$ vanishes identically on
$\nu.$ Because $\phi$ is the difference of two solutions having
identical boundary values, $\phi\left(1,\pm\theta_0\right)=0.$
Also, $\phi_r$ is zero along the lines $\theta=\pm\theta_0.$ Then
$\chi_r=\psi_2=-2\phi_r\phi_\theta=0$ on these lines. Integrating,
$\chi\left(r,-\theta_0\right)=c_1$ and
$\chi\left(r,\theta_0\right)=c_2,$ where $c_1$ and $c_2$ are
constants. On the arc $\nu,$
\begin{equation}\label{sonic}
    \chi_\theta = \psi_1=-\phi_\theta^2\leq 0
\end{equation}
so $c_1\geq c_2.$ On the intersection of the line $r=\varepsilon$
with $\partial\Omega,$
\[
\chi_\theta =
\psi_1=\varepsilon\left(1-\varepsilon^2\right)\phi_r^2\geq 0,
\]
as $\phi_\theta$ is zero there by the homogeneous boundary
conditions, implying that $c_2\geq c_1.$ We conclude that
$c_1=c_2.$ Now (\ref{sonic}) implies that $\chi_\theta\equiv 0$ on
$\nu.$ Integrating the differential equation
$\chi_\theta=-\phi_\theta^2=0$ and using the boundary conditions
$\phi\left(1,\pm\theta_0\right)=0,$ we conclude that $\phi$
vanishes identically on the parabolic line $\nu.$ The homogeneous
boundary conditions now imply that $\phi$ vanishes on all of
$\partial\Omega^+.$ The maximum principle for nonuniformly
elliptic equations implies that $\phi$ vanishes identically on
$\Omega^+$ (see the Remark following the proof of Theorem 3.1 in
Sec.\ 3.1 of \cite{GT}).

Denote by $\Gamma$ the set of polar lines of $\gamma,$ where
$\gamma$ is the set of chords $x=\tau$ for $x_0<\tau<1.$ Then
$\Gamma$ is a set of characteristic lines to eq.\ (\ref{pol3}),
which are intersecting pairs of tangent lines to the unit circle.
Order $\Gamma$ by decreasing radial distance, from the origin, of
the point of intersection for the two polar lines of $\tau.$ Then
the indexed set $\Gamma_\lambda,$ beginning with the polar lines
of $\gamma_0$ and having a limit point at the vertical line $x=1,$
foliates the subdomain $\Omega^-$ in the sense that $\Omega^-$ can
be expressed as the uncountable union of the elements of
$\left\lbrace \Gamma_\lambda\right\rbrace.$

On elements of $\Gamma_\lambda$ we have the characteristic
equation
\begin{equation}\label{stokes1}
    dr^2 - r^2\left(r^2-1\right)d\theta^2=0,
\end{equation}
so
\[
d\chi=\chi_\theta d\theta+\chi_r dr = \left(\chi_\theta\pm\chi_r
r\sqrt{r^2-1}\right)d\theta
\]
and, using (\ref{psi1}) and (\ref{psi2}),
\[
\frac{d\chi}{d\theta}=r^2\left(1-r^2\right)\phi_r^2-\phi_\theta^2\mp
2r\sqrt{r^2-1}\phi_r\phi_\theta
\]

\[
=-\left(r\sqrt{r^2-1}\phi_r\pm\phi_\theta\right)^2\leq 0.
\]
Because $\chi_\theta=\chi_r=0$ on $\nu,$ we know that $\chi$ is
constant on $\nu.$ Integrating along elements of $\Gamma_\lambda$
and using the constancy of $\chi$ on the arc $\nu$ now implies
that $\chi$ is constant, and thus $\chi_\theta$ vanishes, on all
of $\Omega^-.$ But $\chi_\theta=\psi_1$ and $1-r^2<0$ on
$\Omega^-,$ so (\ref{psi1}) implies that
$\phi_r^2=\phi_\theta^2=0$ on $\Omega^-.$ Thus $\phi$ is constant
on $\Omega^-.$ Because $\phi$ vanishes on $\nu,$ we find by
continuity that the constant is zero.

This completes the proof of Theorem 1.

\bigskip

\textbf{Corollary 2}. \emph{Considering (\ref{pol3}) as an
equation on a domain $\Omega$ of the Cartesian $r\theta$-plane,
the closed Dirichlet problem for classical solutions of eq.\
(\ref{pol3}) is over-determined on $\Omega.$}

\bigskip

\textbf{Remarks}. \emph{i)} A nonexistence result of an entirely
different kind, for solutions of a semilinear elliptic-hyperbolic
equation, is presented in \cite{LP1}. That result is based on an
elliptic-hyperbolic extension of the Pohozaev identities of
elliptic theory.

\emph{ii}) Defining the first-order operator $N\phi=-2\phi_r$ and
the 1-form
\[
\psi = \psi_1d\theta+\psi_2dr,
\]
eq.\ (\ref{multipl}) can be interpreted formally as associating a
conservation law to the second-order operator $L\phi$ defined by
eq.\ (\ref{pol3}) via the method of multipliers:
\[
0=\int\int N\phi\cdot L\phi \,drd\theta = \int\int d\psi.
\]
(Compare this equation with p.\ 262 of \cite{LP2} and eqs.\ (5),
(6) of \cite{M2}; see also the equation preceding eq.\ (7.1) of
\cite{ES}, taking $\partial U/\partial t =0$ in that equation.)
However, this interpretation is only applicable to a sufficiently
small coordinate patch about an arc of the sonic curve. In
general, $\chi$ is only incompletely specified by eq.\
(\ref{multipl}); however, in the case considered in Theorem 1,
regions on which $\phi$ must vanish are deduced from boundary
conditions and the particular geometry of the domain.

\emph{iii}) Despite the close similarity of Theorem 1 to Sec.\ 3
of \cite{MSW}, eq.\ (\ref{pol3}) differs from the corresponding
equation, (23), of \cite{MSW} in its type-change function and
lower-order terms. Moreover, because the hyperbolic boundary
depends on the form of the characteristic equation, this
difference in type-change function affects the geometry of the
domain.

\emph{iv}) The characteristic equation (\ref{stokes1}), in its
Cartesian form
\begin{equation}\label{stokes2}
    \left(a^2-x^2\right)dy^2+2xy\,dx\,dy+\left(a^2-y^2\right)dx^2=0
\end{equation}
for $a$ a constant, seems to have been introduced in 1854 by G. G.
Stokes, in the same Cambridge Smith's Prize examination in which
Stokes' Theorem first appeared (\emph{c.f.} \cite{St} and eq.\ 1
of \cite{O10}). Although no geometric context was suggested in the
examination question, Stokes drew the attention of examinees to
the possibility of singular solutions.

\subsection{A strongly well-posed boundary-value problem}

Solutions which exist in the closure of the graph of the
differential operator are said to be \emph{strong;} \emph{c.f.}
\cite{Fr}, p.\ 354. Precisely, a \emph{strong solution} of a
boundary-value problem for an operator equation $Lu=f,$ with $f\in
L^2,$ is an element $u\in L^2$ for which there exists a sequence
$u^{\nu }$ of continuously differentiable functions, satisfying
the boundary conditions, for which
\[
\lim_{\nu \rightarrow \infty }\left\| u^{\nu }-u\right\|_{L^2} =
\lim_{\nu \rightarrow \infty }\left\| Lu^{\nu }-f\right\|_{L^2}
=0.
\]
A consequence of this definition is the uniqueness of strong
solutions.

A system of two differential equations on a domain of
$\mathbb{R}^2$ is \emph{symmetric positive} \cite{Fr} if it can be
written in the form of the matrix equation
\begin{equation}\label{spos}
    Lw\equiv A^1w_x+A^2w_y+Bw=\varphi,
\end{equation}
where $w=\left(w_1,w_2\right)$ is a vector, the matrices $A^1$ and
$A^2$ are symmetric, and the symmetric part $\kappa^\ast$ of the
matrix $\kappa \equiv B -(1/2)\left( A_x^1 + A_y^2\right)$ is
positive definite. Here and below, an asterisk denotes
symmetrization: $\kappa^\ast = (\kappa+\kappa^T)/2.$

Considering a solution $\omega$ to (\ref{spos}) as a mapping from
the domain $\Omega$ to a finite-dimensional vector space $V,$
denote by $\partial V$ the restriction of $V$ to the boundary
$\partial\Omega$ of $\Omega.$ We will use in a fundamental way a
theorem by Friedrichs \cite{Fr}; see also \cite{LaP} and
\cite{Sa}.

\bigskip

\textbf{Friedrichs' Theorem on Symmetric Positive Systems}.
\emph{Let $\Omega$ be a bounded domain of $\mathbb{R}^2$ having
$C^2$ boundary $\partial \Omega.$ Consider the matrix $\beta =
n_1A_{|\partial \Omega}^1 + n_2A_{|\partial \Omega}^2,$ where
$n=\left(n_1,n_2\right)$ is the outward-pointing normal vector to
$\partial \Omega.$ Suppose that the matrix $\beta$ admits a
decomposition $\beta = \beta_++\beta_-,$ for which the sum of the
null spaces for $\beta_+$ and $\beta_-$ spans $\partial V;$
$\mathfrak{R}_+\cap \mathfrak{R}_-=0,$ where $\mathfrak{R}_\pm$ are
the ranges of $\beta_\pm;$ the matrix $\mu=\beta_+-\beta_-$
satisfies $\mu^\ast \geq 0.$ Then the boundary-value problem given
by}
\begin{equation}\label{Fried}
    Lw=\varphi \mbox{ in } \Omega,
\end{equation}
\[
\beta_-w=0 \mbox{ on } \partial\Omega,
\]
\emph{where $L$ is given by (\ref{spos}), has a strong solution
whenever the system is symmetric positive and the components of
$\varphi$ are square-integrable.}

\bigskip

A boundary-value problem which possesses a strong solution is said
to be \emph{strongly well-posed}. We show the existence of such a
problem for a class of equations of Keldysh type. We initially
consider equations having the form
\begin{equation}\label{div}
    \left[K(\eta)u_\eta\right]_\eta+u_{\xi\xi}+k u_\xi=f\left(\eta,\xi\right),
\end{equation}
where $K'(\eta)>0,$ $k$ is a nonzero constant, $K(\eta)<0$ for
$0\leq\eta<\eta_{crit}$ and $K(\eta)>0$ for $\eta_{crit}<\eta\leq
R.$ We do not expect classical solutions, so rather than study
this equation directly, we consider the associated system
\begin{equation}\label{mperturb}
    L w=A^1 w_\eta+A^2 w_\xi+B w= F,
\end{equation}
where $L$ is a first-order operator, $ w=\left( w_1(\eta,\xi),
w_2(\eta,\xi)\right),$ $F=\left( f,0\right),$

\begin{equation}\label{ma1}
    A^1=\left(%
\begin{array}{cc}
  K(\eta) & 0 \\
  0 & -1 \\
\end{array}%
\right), \,A^2=\left(%
\begin{array}{cc}
  0 & 1 \\
  1 & 0 \\
\end{array}%
\right),
\end{equation}
and
\begin{equation}\label{ma2}
    B=\left(%
\begin{array}{cc}
  K'(\eta) & k \\
  0 & 0 \\
\end{array}%
\right).
\end{equation}

We interpret $\eta$ as a radial coordinate and $\xi$ as an angular
coordinate, so that the system (\ref{mperturb})-(\ref{ma2}) is
defined on a closed disc of radius $R.$ The system is equivalent
to (\ref{div}) if the components of $w$ are $C^2$ and
$w_1=u_\eta,$ $w_2=u_\xi.$

\bigskip

\textbf{Theorem 3}. \emph{Suppose that there is a positive
constant $\nu_0$ such that $K'(\eta)\geq \nu_0.$ Let there be
continuous functions $\sigma(\xi)$ and $\tau(\xi)$ such that the
boundary condition}
\begin{equation}\label{bndry}
    \sigma(\xi)w_1+\tau(\xi)w_2=0,
\end{equation}
\emph{where the product $\sigma(\xi)\tau(\xi)$ is either strictly
positive or strictly negative and has sign opposite to the sign of
$k.$ Then the boundary-value problem
(\ref{mperturb})-(\ref{bndry}), with $K$ and $k$ as defined in
eq.\ (\ref{div}), possesses a strong solution on the closed disc
$\left\lbrace \left(\eta,\xi\right)|0\leq\eta\leq R\right\rbrace$
provided $|K(0)|$ is sufficiently small.}

\bigskip

\emph{Proof}. Multiply the terms of eq.\ (\ref{mperturb}) by the
matrix
\[
E=\left(%
\begin{array}{cc}
  a & -cK(\eta) \\
  c & a \\
\end{array}%
\right),
\]
where $a$ and $c$ are constants; the sign of $c$ is chosen so that
$\sigma\tau c<0$ (so $ck>0$); $a>0;$ and $|c|$ is large. The
matrix $E$ is nonsingular provided
\[
\det{E} = a^2+c^2K(\eta)\ne 0.
\]
Because $K(\eta)$ is continuous, increasing, and
$K(\eta_{crit})=0,$ the invertibility condition for $E$ becomes
\[
\min_{\eta\in[0,R]}|K(\eta)|<\frac{a^2}{c^2}.
\]
This condition will be satisfied provided $|K(0)|$ is sufficiently
small.

The symmetric part $\kappa^\ast$ of the matrix
\[
\kappa = EB-(1/2)\left[(EA^1)_\eta + (EA^2)_\xi\right]
\]
has determinant
\[
\Delta=\frac{ak}{2}\left[cK'(\eta)-\frac{ak}{2}\right]\geq
\frac{ak}{2}\left[c\nu_0-\frac{ak}{2}\right],
\]
so the resulting system is symmetric positive provided $|c|$ is
sufficiently large. The proof will be complete once we show that
the boundary conditions are admissible. Proceeding as in
\cite{To}, choose the outward-pointing normal vector to have the
form $n=K^{-1}(\eta)d\eta.$

Then on the boundary $\eta=R,$
\[
\beta=\left(%
\begin{array}{cc}
  a & c \\
  c & -aK^{-1}(R) \\
\end{array}%
\right).
\]
Choose
\[
\beta_{-}=
\]
\[
\frac{1}{\sigma^2+\tau^2}\left(%
\begin{array}{cc}
  \sigma\tau c+\sigma^2 a & \tau^2c+\sigma\tau a \\
  -\sigma \tau aK^{-1}(R)+\sigma^2c & -\tau^2aK^{-1}(R)+\sigma\tau c \\
\end{array}%
\right).
\]
Choose $\beta_{+}=\beta-\beta_{-}.$ Then $\beta_{-}w=0,$ as
(\ref{bndry}) implies that $w_2=-(\sigma/\tau)w_1$ on the circle
$\eta=R.$ Moreover,
\[
\mu =
\]
\[
\frac{1}{\sigma^2+\tau^2}\left(%
\begin{array}{cc}
  \left(\tau^2-\sigma^2\right)a-2\sigma\tau c & \left(\sigma^2-\tau^2\right)c-2\sigma\tau a \\
  \left(\tau^2-\sigma^2\right)c+2\sigma\tau |aK^{-1}(R)| & \left(\tau^2-\sigma^2\right)aK^{-1}(R) -2\sigma\tau c \\
\end{array}%
\right),
\]
implying that
\[
\mu^\ast =
\]
\[
\frac{1}{\sigma^2+\tau^2}\left(%
\begin{array}{cc}
  \left(\tau^2-\sigma^2\right)a-2\sigma\tau c & \sigma\tau a\left( K^{-1}(R)-1\right)\\
  \sigma\tau a\left( K^{-1}(R)-1\right) & \left(\tau^2-\sigma^2\right)aK^{-1}(R) -2\sigma\tau c \\
\end{array}%
\right).
\]
If $\sigma\tau<0,$ choose $c>0;$ if $\sigma\tau>0,$ choose $c<0.$
Then the matrix $\mu^\ast$ will be non-negative provided $|c|$ is
sufficiently large.

Now
\[
\mathfrak{R}_{-}=\frac{\sigma \omega_1 +\tau \omega_2}{\sigma^2+\tau^2}\left(%
\begin{array}{c}
  \tau c + \sigma a \\
  -\tau aK^{-1}(R) +\sigma c \\
\end{array}%
\right)
\]
and
\[
\mathfrak{R}_{+}=\frac{\tau \omega_1 -\sigma \omega_2}{\sigma^2+\tau^2}\left(%
\begin{array}{c}
  \tau a - \sigma c \\
  \sigma aK^{-1}(R) +\tau c \\
\end{array}%
\right),
\]
so $\mathfrak{R}_{^-} \cap\mathfrak{R}_{^+}=0.$ Because conditions
are given on the entire boundary of the disk, the null space of
$\beta_-$ alone spans the range of $\partial V.$

The invertibility of $E$ completes the proof of Theorem 3.

\bigskip

\textbf{Remarks}. \emph{i}) Equation (\ref{div}) is a lower-order
perturbation of an equation studied by Magnanini and Talenti in
the context of singular optics \cite{MaTa}. Those authors
considered $L^2$ data prescribed on the boundary of the unit disc.
They were able to construct an explicit, smooth solution in the
interior of the disc having a point singularity at the origin.
Moreover, they showed the problem to be weakly well-posed on the
entire domain. Theorem 3 of this section replaces Theorem 3 of
\cite{O10} which claims, on the basis of an erroneous proof
\cite{O10e}, that a strong solution to a closed boundary-value
problem exists for an arbitrarily small lower-order perturbation
of the equation studied in \cite{MaTa}. In the present theorem the
perturbation is lower-order but not arbitrarily small.

\emph{ii}) A similar result has been proven by Torre \cite{To} for
the equation
\begin{equation}\label{torre}
    \frac{1}{\eta}\left(\eta
    u_\eta\right)_\eta+\left(\frac{1}{\eta^2}-\tilde\omega^2\right)u_{\xi\xi}=f\left(\eta,\xi\right),
\end{equation}
where $f$ is a sufficiently smooth function and $\tilde\omega$ is
a constant. In \cite{To}, condition (\ref{bndry}) is imposed on
functions $\sigma,$ $\tau$ such that the product $\sigma\tau$ does
not vanish on the outer boundary of an annulus and $\sigma,$
$\tau$ satisfy the Dirichlet conditions $\sigma=1,$ $\tau=0$ on
the inner boundary. Because eq.\ (\ref{torre}) is a helically
reduced wave equation in $\mathbb{M}^{2,1},$ it provides a toy
model for the reduction of the Einstein equations by a helical
Killing field (\emph{c.f.} Examples 1 and 4 of Sec.\ 6, below). In
distinction to eq.\ (\ref{div}), which is of Keldysh type, eq.\
(\ref{torre}) is of Tricomi type.

\bigskip

This section has been devoted to closed boundary-value problems
for harmonic fields on extended $\mathbb{P}^2$ and similar
equations of Keldysh type. The vast literature on open
boundary-value problems for equations of the form (\ref{t-type})
and its nonlinear relatives has been ignored. The literature on
such problems published during the first half of the twentieth
century is reviewed in \cite{Be}; for more recent literature, see
\cite{Cp}. See also \cite{Mw}. All three references concentrate on
the literature in mathematical fluid dynamics, but results also
exist on boundary-value problems for nonlinear equations related
to (\ref{t-type}) that arise in connection with the isometric
embedding of Riemannian manifolds into a higher-dimensional
Euclidean space \cite{Lin},\cite{NM}, \cite{Z}.

\section{Duality}

In this and the following section we return to the geometric
variational problems introduced in Section 2. But we will
eventually replace the linear Hodge theory of eqs.\ (\ref{hodge})
by a nonlinear Hodge theory based on an elliptic-hyperbolic
extension of the Yang-Mills equations.

\subsection{Hodge duality of surfaces}

It has been known for nearly a century that every solution of the
Euclidean minimal surface equation
\begin{equation}\label{M}
    \nabla\cdot\left(\frac{\nabla u}{\sqrt{1+|\nabla u|^2}}\right)=0
\end{equation}
over all of $\mathbb{R}^2$ is an affine linear function \cite{Bn}.
Geometrically, this means that a non-parametric minimal surface of
$\mathbb{R}^3$ is a plane. This \emph{Bernstein property} was
extended to entire solutions of the Lorentzian maximal space-like
hypersurface equation
\begin{equation}\label{L}
    \nabla\cdot\left(\frac{\nabla u}{\sqrt{1-|\nabla u|^2}}\right)=0
\end{equation}
by Calabi \cite{Ca}: a maximal space-like hypersurface of
$\mathbb{M}^{2,1}$ is a space-like plane. Calabi's result was
later extended to space-like hypersurfaces of $n$-dimensional
Minkowski space by Cheng and Yau \cite{CY}.

Yang has interpreted this kind of equivalence between equations
(\ref{M}) and (\ref{L}) in the language of differential forms
(\cite{Y}, eqs.\ (2.23)$-$(2.29); see also \cite{AP}). In Yang's
interpretation, Calabi's equivalence becomes duality under the
Hodge isomorphism.

Yang's argument involves writing eq.\ (\ref{quasi}) in the dual
form \cite{SS1}
\begin{equation}\label{HM}
    \delta\left[\rho(Q)\omega\right]=d\omega=0,
\end{equation}
where, as in (\ref{hodge}), $d:\Lambda^p\rightarrow\Lambda^{p+1}$
is the exterior derivative with formal adjoint
$\delta:\Lambda^p\rightarrow\Lambda^{p-1};$
\[
    Q(\omega)=|\omega|^2=*\left(\omega\wedge*\omega\right);
\]
$*:\Lambda^p\rightarrow\Lambda^{n-p}$ is the Hodge involution. If
the surface is embedded in Euclidean 3-space, then we choose
\begin{equation}\label{den}
    \rho(Q)=\frac{1}{\sqrt{1+Q}}.
\end{equation}
If the surface is embedded in Minkowski 3-space, then
\begin{equation}\label{den1}
    \rho(Q)=\frac{1}{\sqrt{|1-Q|}}.
\end{equation}
In the latter case it is necessary to distinguish space-like
surfaces, for which $Q<1,$ from time-like surfaces, for which
$Q>1.$ The surface $Q=1$ is the light cone. Note that none of
these surfaces can be found in $\mathbb{M}^{2,1}$ without solving
the system (\ref{HM}), (\ref{den1}).

If $\omega$ is a $p$-form satisfying (\ref{HM}) over a region
having trivial de Rham cohomology, then the $n-p-1$-form
$*\left[\rho(Q)d\omega\right]$ is closed, and thus is the
differential of an $n-p-2$-form $\sigma:$
\[
d\sigma=*\left[\rho(Q)d\omega\right].
\]
If $\rho(Q)$ is given by (\ref{den}), then this implies that
\[
0=d^2\omega=d\left[(-1)^{p(n-p)+n-1}*\frac{d\sigma}{\sqrt{1-|d\sigma|^2}}\right]=\pm\delta\left[\tilde\rho(Q)d\sigma\right],
\]
with $\tilde\rho(Q)$ given by (\ref{den1}); see eqs.\ (2.23-2.30)
of \cite{Y} for details.

Yang obtains from this duality, not the Bernstein property, but
rather a \emph{Liouville property} (\cite{Y}, eq. (2.31)): if either
$d\omega$ or $d\sigma$ have finite energy on $\mathbb{R}^n$ in the
sense that
\begin{equation}\label{ef}
    E=\int_{\mathbb{R}^n}\int_0^Q\rho(s)ds*1<\infty
\end{equation}
for $\rho(Q)$ given by either (\ref{den}) or (\ref{den1}), then
\begin{equation}\label{bern}
    d\omega=0 \mbox{ or } d\sigma=0.
\end{equation}
See Secs.\ 1-4 of \cite{SSY} for more details of this and similar
Liouville theorems. We return to this topic in Sec.\ 5.

\subsection{Hodge duality of flow metrics}

It is also possible to establish a kind of Hodge duality between
extremal surfaces in $\mathbb{M}^{2,1}$ and a fully quasilinear
model for compressible potential flow. If $L$ is the operator of
eq.\ (\ref{secord}), we define as in \cite{Be} the \emph{flow
metric} for the operator $L$ to be the metric tensor $g_{ij}$
having distance element
\[
ds_{L}^{2}\equiv \alpha\left( x,y\right) dy^{2}-2\beta \left(
x,y\right) dxdy+\gamma \left( x,y\right) dx^{2}.
\]
The metric $g_{ij}$ is Riemannian for regions on which $L$ is an
elliptic operator and Lorentzian for regions on which $L$ is a
hyperbolic operator.

This idea does not use the linearity of $L$ in any essential way,
and its original motivation seems to have been the quasilinear
equation for the steady flow of an ideal gas. Define the quantity
\[
c^2=1-\frac{\gamma-1}{2}\left(u^2+v^2\right),
\]
where $u$ and $v$ are the components of the velocity of an ideal
compressible fluid in the $x$ and $y$ directions, respectively;
$\gamma>1$ is the adiabatic constant of the medium. (The notation,
which is historical, implies that the flow metric has
traditionally not been applied beyond the cavitation velocity
$u^2+v^2=2/\left(\gamma-1\right).$ The gas is assumed to be
adiabatic and isentropic in order to avoid having to specify
thermodynamic variables.)

In the irrotational case we can define a potential function
$\psi\left(x,y\right)$ for which
\[
d\psi = udx + vdy
\]
and
\[
*d\psi = udy-vdx,
\]
where the centered asterisk again denotes Hodge involution. Then
the flow metric for the continuity equation of compressible ideal
flow in the plane,
\[
\left(c^2-u^2\right)u_x-2uvu_y+\left(c^2-v^2\right)v_y=0,
\]
has distance element
\begin{equation}\label{fl1}
    ds^2=c^2\left(dx^2+dy^2\right)-\left(*d\psi\right)^2;
\end{equation}
the flow metric of the classical minimal surface equation
\begin{equation}\label{min}
    \left(1+v^2\right)u_x-2uvu_y+\left(1+u^2\right)v_y=0
\end{equation}
for the graph of a smooth function $z=\psi\left(x,y\right)$ has
distance element
\begin{equation}\label{fl2}
    ds'^2=dx^2+dy^2+d\psi^2.
\end{equation}

Each of the two flow metrics (\ref{fl1}) and (\ref{fl2})
decomposes into a conformally Euclidean part and a non-Euclidean
part. Moreover, if the classical minimal surface equation
(\ref{min}) for smooth surfaces embedded in $\mathbb{R}^3$ is
replaced by the equation (\ref{quasi}) for surfaces embedded in
$\mathbb{M}^{2,1},$ then the non-Euclidean part of that flow
metric is the Hodge dual of the non-Euclidean part of the flow
metric for ideal compressible fluids: the flow metric for
(\ref{quasi}) has element of distance $ds''$ given by
\[
ds''^2=dx^2+dy^2-d\psi^2.
\]

For more details on duality of flow metrics, see Sec.\ 2.1 of
\cite{O6}.

The approach to duality initiated by eq.\ (\ref{HM}) seems to have
been introduced in \cite{SS1}, in the context of compressible flow
on a compact Riemannian manifold; also compare \cite{ISS} and
\cite{MP} with \cite{SS2}.

\section{Gauge invariance}

\subsection{Twisted nonlinear Hodge equations}

It is natural to ask why the Bernstein property for extremal
surfaces in \cite{Bn}, \cite{Ca}, and \cite{CY} becomes a
Liouville property for differential forms in eq.\ (\ref{bern}).
The Bernstein property for solutions of the two equations
(\ref{M}) and (\ref{L}) has an intuitive interpretation because
these partial differential equations can be embedded in a
geometric theory, the theory of extremal hypersurfaces. In order
to obtain an intuitive interpretation of Yang's Liouville property
for solutions of the two systems (\ref{HM}), (\ref{den}) and
(\ref{HM}), (\ref{den1}), it would be useful to embed those
partial differential equations as well in an appropriate geometric
theory. Yang observes \cite{Y} that we can do so if the order of
the differential form $\omega$ is 1 or $n-1.$ We can also do so in
the case studied by Yang, in which the order of the differential
form is 2.

In that case eq.\ (\ref{HM}), with $\rho$ given by (\ref{den1}),
is the Born-Infeld equation for an electromagnetic field $\omega$
\cite{BI}, \cite{Y}. There is a generalization of the Born-Infeld
equations which illuminates their geometric properties and, in
particular, gives a geometric interpretation of the Liouville
property for those equations.

Proceeding as in \cite{O4} and \cite{O5}, we denote by $X$ a
vector bundle having compact structure group $G\subset SO(m)$ and
define $X$ over a smooth, finite, oriented, \textit{n}-dimensional
Riemannian manifold $M.$ Form an admissible class of connections
by choosing a smooth base connection $D$ in the space of
connections compatible with $G$ and considering the class of
connections $D+A,$ where $A$ is a section of $ ad\,X\otimes
T^{*}M$ which lies in the largest Sobolev space for which the
energy functional $E$ given by (\ref{ef}) is finite (with the
domain $\mathbb{R}^n$ replaced by $M$); details are given in
\cite{U} for the case $\rho \equiv 1$. Denote by $F_A$ the
curvature 2-form associated to a connection 1-form $A.$ Then
$Q=|F_{A}|^{2}=\left \langle F_{A},F_{A}\right\rangle $ is an
inner product on the fibers of the bundle $ad\,X\otimes \Lambda
^{2}\left( T\;^{\ast }M\right) .$ The inner product on $ad\,X$ is
induced by the normalized trace inner product on $ SO(m) $ and the
inner product on $\Lambda ^{2}\left( T\;^{\ast }M\right)$ is
induced by the exterior product $\ast \left( F_{A}\wedge \ast
F_{A}\right).$ Sections of the automorphism bundle $Aut\,X$ are
\emph{gauge transformations}. These maps act tensorially on
$F_{A}$ but affinely on $ A $; see, \textit{e.g.,} Sec.\ 6.4 of
\cite{MM}. Note that it is possible to describe the geometry of
$F_A$ either in terms of a principal bundle or in terms of the
associated vector bundle. The choice is a matter of convenience;
see, \emph{e.g.}, Sec.\ 2.3 of \cite{MM} for a discussion.

In the smooth case we take variations by computing
$(d/dt)(F_{D+tA})$ at the origin of $t.$ Using the fact that for
any smooth section $\sigma $ we have (\cite{O4}, Sec.\ 1)
\[
F_{D+tA}(\sigma )=\left( F+tDA+t^2A\wedge A\right) (\sigma ),
\]
we obtain
\[
\delta E=\frac{1}{2}\int_M\rho (Q)\delta Q\;dM
\]

\begin{equation}\label{vari}
=\frac{1}{2}\int_M\rho (Q)\frac d{dt}_{|t=0}|F+tDA+t^2A\wedge
A|^2\;dM.
\end{equation}
Letting $t=0,$ the right-hand side of (\ref{vari}) can be written
\begin{equation}\label{euler}
\int_M\rho (Q)\,\langle DA,\,F_A\rangle \;dM=\int_M\langle
DA,\,\rho (Q)F_A\rangle \;dM.
\end{equation}
We assume that either $\partial M=0$ or, if not, that $F_A$
satisfies a ``Neumann'' boundary condition of the form
\begin{equation}\label{neum}
i^{*}(*F)=0
\end{equation}
on $\partial M,$ where $i^{*}$ is the pull-back under inclusion of
the boundary of $M$ in $M$. This is equivalent in local
coordinates to prescribing zero boundary data for $F$ in a
direction normal to $\partial M;$ see \cite{Ma} for details in the
case $\rho \equiv 1.$

Set $\delta E$ equal to zero. Then (\ref{vari}) and (\ref{euler})
imply
\[
0=\int_M\left\langle DA,\,\rho (Q)F_A\right\rangle \;dM=
\]
\[
\int_{\partial M}A_\vartheta \wedge \left( \,\rho (Q)F_A\right)
_N\,+\int_M\left\langle A,\,D^{*}\left( \rho (Q)F_A\right)
\right\rangle \;dM,
\]
where $D^{*}$ denotes the formal adjoint of the exterior covariant
derivative $D;$ $\vartheta $ denotes tangential component on the
boundary and $N$, the normal component there. Condition
(\ref{neum}) implies Euler-Lagrange equations of the form
\cite{O2}, \cite{O3}
\begin{equation}\label{vari1}
D^{*}\left( \rho (Q)F\right) =0.
\end{equation}
Because $F$ is a curvature 2-form, it satisfies an additional
condition
\begin{equation}\label{bian}
DF=0,
\end{equation}
the second Bianchi identity.

If $\rho\equiv 1,$ then eqs.\ (\ref{vari1}), (\ref{bian})
degenerate to the Euclidean Yang-Mills equations of high-energy
physics. Viewed in another way, one obtains eqs.\ (\ref{vari1}),
(\ref{bian}) from eqs.\ (\ref{HM}) by twisting the cotangent
bundle in which solutions of (\ref{HM}) live. This twist is
represented analytically by a nonvanishing Lie bracket with $A.$
Precisely, $A\in \Gamma \left( M,ad\,X\otimes T^{\ast }M\right) $
is a connection 1-form on $X$ having curvature 2-form
\[
F_{A}=dA+\frac{1}{2}\left[ A,\,A\right] =dA+A\wedge A,
\]
where [\ ,\ ] is the bracket of the Lie algebra $\Im ,$ the fiber
of the adjoint bundle $ad\,X.$ Thus eqs.\ (\ref{vari1}),
(\ref{bian}) can be characterized algebraically as a
non-commutative version of (\ref{HM}), or geometrically as
\emph{twisted nonlinear Hodge equations}.

Equations (\ref{vari1}), (\ref{bian}) change from elliptic to
hyperbolic type along the curve
\[
\frac{d}{dQ}\left(Q\rho^2(Q)\right)=0.
\]
So, in this case as well, the elliptic and hyperbolic regions of
the equations cannot be found without solving the system. For
$\rho$ given by eq.\ (\ref{den}), eqs.\ (\ref{vari1}),
(\ref{bian}) do not change type, but ellipticity degenerates as
$Q$ tends to infinity.

Because eqs.\ (\ref{HM}) are equivalent to eqs.\ (\ref{vari1}),
(\ref{bian}) in the special case of an abelian gauge group $G,$
the closed 2-form $\omega$ of the classical Born-Infeld theory is,
geometrically, an abelian special case of the curvature 2-form
$F_A$ (\emph{c.f.} \cite{O5}, Sec.\ 1). We can interpret Yang's
Liouville theorem given by eq.\ (\ref{bern}) for solutions of the
system (\ref{HM}) with (\ref{den}) or (\ref{den1}), in the
generalized perspective of \cite{O3}-\cite{O5}, as the assertion
that a finite-energy solution of (\ref{HM}) on $\mathbb{R}^n$ with
density (\ref{den}) or (\ref{den1}) is associated with a bundle
having zero curvature, just as entire solutions in $\mathbb{R}^3$
or $\mathbb{M}^{2,1}$ are associated with surfaces having zero
curvature. That geometric interpretation is reflected in the
degeneration of the Bernstein property of \cite{Ca}, for entire
solutions of (\ref{M}) and (\ref{L}), to a Liouville property in
\cite{Y} for entire solutions of (\ref{HM}) with (\ref{den}) or
(\ref{den1}).

Liouville theorems have also been derived for solutions of
(\ref{vari1}) and similar geometric objects; see, \emph{e.g.,}
\cite{O1}, \cite{O3}, Sec.\ 5 of \cite{O8}, and Secs.\ 3 and 4 of
\cite{SSY}. An $n$-dimensional Born-Infeld system related to
time-like extremal surfaces in higher-dimensional Minkowski space
is studied in \cite{Ko}.

In \cite{Ka} a physical model for compressible ideal flow with
$SO(3)$ gauge invariance is proposed. The mathematical structure
of this model is similar in some ways to the variational theory
presented here.

Choosing the function $\rho(Q)$ in eqs.\ (\ref{vari1}) to be
$1+4\gamma Q^{-1},$ where $\gamma$ is a parameter, the gauge group
to be the abelian group $U(1),$ and the underlying metric to be
Robertson-Walker, eqs.\ (\ref{vari1}) become identical to eqs.\
(12) of \cite{NBS}, a nonlinear electrodynamic model for the
accelerated expansion of the universe.

A version of eqs.\ (\ref{vari1}), (\ref{bian}) for mappings exists
\cite{O5}, \cite{O9}, \cite{O11} which is related to harmonic maps
in exactly the same way that eqs.\ (\ref{vari1}), (\ref{bian}) are
related to Yang-Mills fields. This class of maps is a special case
of the $F$\emph{-harmonic maps} introduced by Ara \cite{Ar}.

\subsection{Global triviality on $\mathbb{R}^n$}

In this section we extend the Liouville theorems cited in the
preceding section to the case of an energy functional for which
the variational equations may be elliptic-hyperbolic and the
density may cavitate, in a pointwise sense, on a set of measure
zero. The bundle connection need not be a pointwise solution of
any equation at all in order to satisfy the hypotheses of this
section.

Consider a principal bundle $\Pi$ over a domain $\Omega$ of
$\mathbb{R}^n$ and a 1-parameter family $\phi^t$ of compactly
supported diffeomorphisms of $\Omega$ for which
$\phi^s\circ\phi^t=\phi^{s+t}$ with $\phi^0$ the identity
transformation. This family of reparametrizations can be lifted to
the principal bundle by parallel transport, with respect to an
arbitrary smooth connection, along the curve $x_s=\phi^s(x)$ from
$x_0=x$ to $x_t=\phi^t(x).$ If
$A\in\Gamma\left(\Omega,\Lambda^1\left(ad\,\Pi\right)\right)$ is a
Lie-algebra-valued connection 1-form, then we define
$A^t=\left(\psi^t\right)^\ast A,$ where $\psi^\ast$ is a lifting
of $\phi^\ast$ to $\Gamma\left(\Omega,\Lambda^1\left(ad\,
\Pi\right)\right).$ (The superscripted asterisk denotes an induced
mapping.) For details of this lifting, see \cite{Pr}. We say that
the connection $A$ is $r$\emph{-stationary} with respect to the
energy functional $E$ given by (\ref{ef}) if
\[
\delta_r E =\frac{d}{dt}_{|t=0}E\left(F^t\right)=
\frac{d}{dt}_{|t=0}E\left(F\left(A^t\right)\right)
\]

\[
=\frac{d}{dt}_{|t=0}\int_{\Omega}\int_0^{\left|F^t\right|^2}\rho(s)ds*1=0,
\]
\emph{c.f.} \cite{Al}. The following theorem provides sufficient
conditions under which $A$ must have zero curvature almost
everywhere if $\Omega=\mathbb{R}^n.$ A slightly more complicated
argument will extend the result to manifolds of constant negative
curvature (\emph{c.f.} \cite{Pr}) and to
$\mathbb{R}^n\backslash\Sigma,$ where $\Sigma$ is a singular set
of prescribed codimension (\emph{c.f.} \cite{O1}).

\bigskip

\textbf{Theorem 4}. \emph{Consider the geometric construction of
the preceding paragraph. Assume that $\rho$ is continuously
differentiable, nonnegative, and weightless under conformal
transformations; that there is a positive number $Q_{crit}$ such
that}

\begin{equation}\label{below}
    e(Q)=\int_0^Q\rho(s)ds>0\,\,\forall \,Q\in
    \left(0,Q_{crit}\right),
\end{equation}
\emph{with $Q(F)=|F_A|^2$ defined as in Sec.\ 5.1; that the
restriction of $E$ to a Euclidean $n$-disc of radius $R$ is
$r$-stationary $\forall R
>0;$ that $\rho'(s)\leq 0\,\forall s \geq 0;$ and that}

\begin{equation}\label{growth}
    E_{|B_R}\leq CR^k
\end{equation}
\emph{as $R$ tends to infinity for a positive constant $C$ and a
sufficiently small (positive) constant $k.$ Then if $Q$ is bounded
above by $Q_{crit},$ $Q\left(F(x)\right)$ must be zero for almost
every $x \in \mathbb{R}^n$ for $n>4.$}

\bigskip

\emph{Proof}. Similar theorems under somewhat different hypotheses
have been proven in \cite{Pr}; see also Secs.\ 2 and 5 of
\cite{O3}, and Sec.\ 5 of \cite{O8}. Assume that the lifting
described earlier has been constructed. Denote by $\psi ^{t}$ the
lift of a 1-parameter family of compactly supported
diffeomorphisms of $\Omega $ such that
\[
\psi ^{s}\circ \psi ^{t}=\psi ^{s+t},\;
\]
and\ $\psi ^{0}=identity.$ Define
\begin{equation}
f\equiv \psi ^{t}(x)=x+t\xi (x)+O(t^{2}),
\end{equation}
where
\[
\xi (x)=\frac{d}{dt}_{|t=0}\psi ^{t}\left( x\right)
\]
is \textit{the variation vector field}, to be chosen. We have

\[
\frac{d}{dt}_{|t=0}F _{ij}(f)df^{i}df^{j}=\frac{d}{dt}_{|t=0}F
_{ij}(f)\frac{\partial f^{i}}{\partial x^{k}}dx^{k}\frac{\partial f^{j}}{%
\partial x^{m}}dx^{m}=
\]
\[
\frac{d}{dt}_{|t=0}F _{ij}(f)\left( \delta _{k}^{i}\delta
_{m}^{j}+\delta _{k}^{i}t\frac{\partial \xi ^{j}}{\partial
x^{m}}+\delta _{m}^{j}t\frac{\partial \xi ^{i}}{\partial
x^{k}}+O(t^{2})\right) dx^{k}dx^{m}.
\]
Make the coordinate transformation $x\rightarrow y,$ where
\[
y=\left( \psi ^{t}\right) ^{-1}(x).
\]
Then

\[
\frac{d}{dt}_{|t=0}F _{ij}(f)df^{i}df^{j}= 2F
_{ij}(x)\frac{\partial \xi ^{i}}{\partial x^{k}}dx^{k}dx^{j}.
\]
If $J$ is the Jacobian of the transformation $x\rightarrow y,$
then
\begin{equation}\label{Jac}
    \frac{d}{dt}_{|t=0}J\left[ \left( \psi ^{t}\right) ^{-1}\right] =\frac{d}{dt}%
_{|t=0}\left| \frac{\partial x}{\partial f}\right| =-div\,\xi .
\end{equation}
By hypothesis, $\forall\, R>1,$
\[
0=\delta _{r}E=\frac{d}{dt}_{|t=0}\int_{B_R }e\left( \left\langle
F^t ,F^t \right\rangle \right) \,J\left[ \left( \psi ^{t}\right)
^{-1}\right] \,\ast 1=
\]
\[
\int_{B_R }e(Q)\frac{d}{dt}_{|t=0}J\left[ \left( \psi ^{t}\right) ^{-1}%
\right] \ast 1+
\]

\begin{equation}\label{vr}
\int_{B_R }e^{\prime }(Q)2\left\langle \frac{d}{dt}_{|t=0}F
_{ij}(f)df^{i}df^{j},F _{\ell m}(f)df^{\ell}df^{m}\right\rangle
\ast 1.
\end{equation}
Substitution of (\ref{Jac}) into (\ref{vr}) yields
\[
\int_{B_R }e(Q)\,div\,\xi \ast 1= 4\int_{B_R }e^{\prime
}(Q)\left\langle F _{ij}(x)\frac{\partial \xi ^{i}}{\partial
x^{\ell }}dx^{\ell}dx^{j},F _{ij}(f)df^{i}df^{j}\right\rangle \ast
1.
\]

Choose an orthonormal basis
\[
\left\{ u_{_{i}}\right\} _{i=1}^{n}=\left\{ \frac{\partial }{\partial r},%
\frac{\partial }{\partial \theta _{2}},\ldots ,\frac{\partial
}{\partial \theta _{n}}\right\}
\]
and let \cite{Pr} $\xi =\eta (r)r\cdot \partial/\partial r,$
where: $r$ is the radial coordinate in a curvilinear system; $\eta
(r)\in C_{0}^{\infty }\left[ 0,1\right] ;$ $\eta ^{\prime }(r)\leq
0;$\ $\eta (r)=v\left( r/\tau \right) =1$ for $r\leq \tau ,$ where
$\tau $ is a number
in the interval $(0,1);$ there is a positive number $\delta $ for which $%
\eta (r)=0$ whenever $r$ exceeds $\tau +\delta .$ \ For this
choice of $\xi , $
\begin{equation}\label{cho}
\int_{B_R }e(Q)\,\left( n\eta +r\eta ^{\prime }\right) \ast
1=4\int_{B_R }Q\rho (Q)\eta \,\ast 1+4\int_{B_R }\rho (Q)r\eta
^{\prime }\left| \frac{\partial }{\partial r}\rfloor F \right|
^{2}\,\ast 1.
\end{equation}
Our hypothesis on the sign of $\rho ^{\prime }$ implies that
\[
Q\rho (Q)=\int_{0}^{Q}\frac{d}{ds}\left( s\rho (s)\right) \,ds
\]

\begin{equation}\label{ine}
    =\int_{0}^{Q}\left[ s\rho ^{\prime }(s)+\rho (s)\right]
\,ds\leq \int_{0}^{Q}\rho (s)\,ds=e\left( Q\right) .
\end{equation}
Substitution of (\ref{ine}) into (\ref{cho}) yields
\[
\int_{B_R }e(Q)\,\left( n\eta -4\eta +r\eta ^{\prime }\right)
\ast 1\leq 4\int_{B_R }\rho (Q)r\eta ^{\prime }\left| \frac{\partial }{\partial r}%
\rfloor F \right| ^{2}\,\ast 1.
\]
By construction
\[
r\eta ^{\prime }(r)=-\tau \frac{\partial }{\partial \tau }v\left( \frac{r}{%
\tau }\right) \leq 0.
\]
This yields
\[
0\leq 4\int_{B_R }\rho (Q)\tau \frac{\partial }{\partial \tau
}v\left( \frac{r}{\tau }\right) \left| \frac{\partial }{\partial
r}\rfloor F \right| ^{2}\,\ast 1\leq \int_{B_R }e(Q)\left[ \left(
4-n\right) \eta +\tau \frac{\partial }{
\partial \tau }v\left( \frac{r}{\tau }\right) \right] \,\ast 1.
\]
As $\delta $ tends to zero we obtain
\[
0\leq \left( 4-n+\tau \frac{\partial }{\partial \tau }\right)
\int_{B_{\tau }}e(Q)\,\ast 1.
\]
Multiply this last inequality by the integrating factor $\tau
^{4-(n+1)}$ and integrate over $\tau $ between $r_{1}$ and
$r_{2}$. We find that
\begin{equation}\label{monot}
    r_{1}^{4-n}E_{|B_{r_{1}}}\leq r_{2}^{4-n}E_{|B_{r_{2}}}.
\end{equation}

We can write the growth condition (\ref{growth}) in the form
\begin{equation}\label{growth1}
r^{4-n}E_{|B_{r}}\leq Cr^{4+k-n},
\end{equation}
where $4+k-n<0$ for sufficiently small $k$. \ The right-hand side
of (\ref{growth1}) tends to zero as $r$ tends to infinity. \ The
left-hand side is nonnegative by construction. \ Thus the
conformal energy $r^{4-n}E_{|B_{r}}$ tends to zero on
$\mathbb{R}^{n}$. \ Because by (\ref{monot}) the conformal energy
is nondecreasing for increasing $r$, we conclude that $E$ is
identically zero on $\mathbb{R}^{n}$. \ The vanishing of the
energy on a ball of infinite radius implies the pointwise
vanishing of $Q$ almost everywhere by inequality (\ref{below}).
This completes the proof.

\bigskip

An exactly analogous result holds in the case of a velocity field
for a steady polytropic flow of an ideal compressible fluid; see
Sec.\ 5 of \cite{O8} for details.

\subsection{Local existence and regularity}

In this section we briefly review the fundamental analysis of
eqs.\ (\ref{vari1}), (\ref{bian}). For details, see \cite{O4}; see
also Sec.\ 2 of \cite{O8}.

The local existence of solutions has been shown in the case
$\rho(Q) = Q^{\left(p-2\right)/2}$ for $p$ exceeding $n/2,$ where
$n$ is the dimension of the domain. In this case the energy
functional is Palais-Smale, and the existence of weak solutions
follows by conventional variational theory. The equations are
nonuniformly elliptic for this choice of $\rho.$ It is possible to
show that, in general, any weak solution is H\"older continuous on
compact subdomains whenever $F_A\in L^p$ for $p>n/2$ and $\rho$ is
chosen in such a way that the system is \emph{uniformly} elliptic.

In the special case of an abelian gauge group, a \textit{weak
solution} of (\ref{vari1}), (\ref{bian}) is any curvature 2-form
$F_A$ for which $ \rho (Q)F_A$ is orthogonal in $L^2$ to the space
of $d$-closed 2-forms $ d\zeta \in L^2(B)$ such that $\zeta \in
\Lambda ^1$ has vanishing tangential data on $\partial B.$ For a
nonabelian gauge group, an obvious extension to inhomogeneous
equations allows a weak solution to be defined by the equation
\begin{equation}\label{weak}
\int_B\left\langle d\zeta ,\rho (Q)F_A\right\rangle
*1=-\int_B\left\langle \zeta ,*\left[ A,*\rho (Q)F_A\right]
\right\rangle *1.
\end{equation}
Here $B$ is a Euclidean $n$-disc, so the natural hypothesis on the
domain is that it can be covered by $n$-discs; for example, assume
that the domain boundary has no cusps.

We briefly outline the proof of regularity: The first step is to
derive a weak subelliptic estimate for $|F|^2,$ using difference
quotients to establish the existence of an $H^{1,2}$-derivative.
The subelliptic estimate allows one to conclude from a limiting
argument that $|F|$ is locally bounded by Morrey's Theorem. Now we
apply a mean-value inequality for Lie-algebra-valued sections,
which is valid in the uniformly elliptic case of the equations
(\ref{vari1}), (\ref{bian}). The mean-value inequality is applied
to points in the solution space, so that the $L^2$-difference of
weak solutions can be compared despite the nonlinearity of $\rho.$
Using this inequality, it is possible to estimate the
$L^2$-difference between a 2-form weakly satisfying eqs.\
(\ref{HM}) and a bounded, weak solution of eqs.\ (\ref{vari1}),
(\ref{bian}). The latter is considered in an exponential gauge,
fixed in a Euclidean $n$-disc $B$ centered at the origin of
coordinates in $\mathbb{R}^n.$ Solutions to eqs.\ (\ref{HM}) have
known regularity, and in an exponential gauge, $A(0)=0$ and
$\forall x\in B$
\[
\left| A(x)\right| \leq \frac 12\left| x\right| \cdot \sup_{\left|
y\right| \leq \left| x\right| }\left| F(y)\right|.
\]
Thus elliptic estimates, and the minimizing property of the
variance by the mean with respect to location parameters,
eventually show that the $L^2$-difference of $F$ and its local
mean value decays sufficiently rapidly as the radius of the
underlying domain shrinks to zero that H\"older continuity can be
obtained from the Campanato Theorem. The final step is to show
that the Campanato estimate is preserved under continuous gauge
transformations in a small \textit{n}-disc centered at a point
close to the origin. This allows one to apply a covering argument
which will extend to the entire domain the estimate that was
obtained in an exponential gauge at the origin.

\subsection{Other symmetry groups}

Although the system (\ref{vari1}), (\ref{bian}) is invariant under
the action of the gauge group $G$ and has the potential to change
from elliptic to hyperbolic type, the system is so poorly
understood that it is hard to know how these two properties affect
solutions. In principle, the system cannot even be said to be
elliptic-hyperbolic, as the type of the equation can only be fixed
modulo gauge transformations. However, the gauge invariance can be
broken and (\ref{vari1}), (\ref{bian}) represented as an
elliptic-hyperbolic system whenever the curvature $F_A$ lies in a
sufficiently high Sobolev space. It is this property, which is
shared with the Yang-Mills equations \cite{U}, that allows one to
formulate sufficient conditions for the existence of solutions,
and their regularity in the uniformly ``subsonic" case, as
described in the preceding section.

The distance element on the Beltrami disc is also invariant under
the action of a gauge group, the group of projective
transformations. This invariance has the consequence that the
parabolic line for the Hodge equations (\ref{P1}), (\ref{P2}) on
extended $\mathbb{P}^2$ is gauge-invariant under perspection to
any conic section. So (\ref{P1}), (\ref{P2}) is an example of an
elliptic-hyperbolic system in which gauge invariance has a
measurable effect on the geometry of the problem. One might try to
interpret a well known model for wave propagation in cold plasma,
in which the parabolic line of an elliptic-hyperbolic system is a
parabola with vertex at the origin of $\mathbb{R}^2$ \cite{MSW},
\cite{PF}, as a fixed gauge of a Hodge system on extended
$\mathbb{P}^2.$

Unfortunately, there would be some problems with interpreting the
cold plasma model in terms of waves on extended $\mathbb{P}^2.$
First, it does not appear to simplify the original problem,
although it may illuminate certain similarities in the analysis of
the two systems; compare \cite{O6}, \cite{O7}, and \cite{O12}.
Second, the projective group consists of non-Euclidean motions,
which are noninertial and thus lacking in obvious physical
meaning, so it is unclear how to interpret the mathematical
relation between the two systems in terms of a satisfying physical
theory. Finally, the gauge groups of Sec.\ 5.1 act ``upstairs'' on
a fiber bundle of states. The symmetry group under which the
Laplace-Beltrami equations are invariant acts ``downstairs'' on
the underlying metric, in the manner of the gauge group of general
relativity. Thus it is reasonable to look for physical analogies
of Sec.\ 2 among theories in which the relevant bundle is soldered
to the base space rather than being related to the base space only
by a projection, as in plasma physics, the Born-Infeld model, and
other electromagnetic theories.

Symmetry groups for differential operators on
Riemannian-Lorentzian metrics other than extended $\mathbb{P}^2$
have been computed \cite{LP2}. In particular, the symmetry group
for operators of Tricomi type which satisfy a pure power law has
been shown to correspond to a group of local conformal
transformations with respect to the underlying metric away from
the metric singularity; moreover, the group extends across the
singular surface on which the metric changes type \cite{Pa3}.

\section{A zoo of mixed Riemannian-Lorentzian metrics}

For the reasons cited in the preceding section, it is reasonable
to look to relativity for physical analogies of Sec.\ 2. We
therefore include a brief survey of signature change in the recent
physics literature.

\medskip

1. \emph{Special relativity}: The wave equation on Minkowski
space-time, in a reference frame rotating with constant angular
velocity $\omega$ with respect to another reference frame, is
expressible in cylindrical coordinates $\left(\rho, \varphi,
z\right)$ as the elliptic-hyperbolic equation \cite{JS}
\[
\frac{1}{\rho}\left(\rho
u_\rho\right)_\rho+\left(\frac{1}{\rho^2}-\omega^2\right)u_{\varphi\varphi}+u_{zz}=0.
\]

\medskip

2. \emph{Quantum cosmology}: These examples arise from the
(controversial) \emph{Hartle-Hawking hypothesis} \cite{HH}, that
the universe might have originated as a manifold having Euclidean
signature and subsequently undergone a transition to a model
having Lorentzian signature across a hypersurface which was
space-like as seen from the Lorentzian side. (Note that certain
metrics which are called \emph{Euclidean} by physicists would be
called \emph{Riemannian} by geometers; see footnote 2 of
\cite{GI}.) Some 2-dimensional variants are \cite{JS}:

\emph{i)} continuous change of signature:
\[
ds^2=-tdt^2+dz^2;
\]

\emph{ii)} discontinuous change of signature:
\[
ds^2=-z^{-1}dt^2+dz^2;
\]

\emph{iii)} continuous change of signature with a curvature
singularity:
\[
ds^2=-zdt^2+dz^2;
\]
see also \cite{DEHM}.

These examples have obvious higher-dimensional analogues.

Alternatively, the metric might change from Lorentzian to Kleinian
signature across the line $z=0:$
\[
ds^2=-dt^2+dx^2+dy^2+zdz^2;
\]
this 4-dimensional model has been studied in, \emph{e.g.},
\cite{A}.

\medskip

In connection with the distinction between examples \emph{ii)} and
\emph{iii)}, we note that operators on Riemannian-Lorentzian
metrics which degenerate rather than blow up at the change of
signature have been studied by mathematicians as well as
physicists \cite{Che2}; see also \cite{Che1} and the discussion in
Sec.\ 5 of \cite{Pa3}.

It is interesting for mathematicians that although many of the
criticisms of Hartle and Hawking's widely discussed proposal
concern physical predictions that it implies, controversy also
arises from mathematical ambiguities in geometric analysis on
mixed Riemannian-Lorentzian metrics and from the variety of
potential singularities of such metrics. See \cite{P} and the
references therein for a recent discussion of the physical
predictions; see, \emph{e.g.}, \cite{DEH}, \cite{DEHM}, \cite{E},
\cite{F}, \cite{H1}, \cite{HD1}, \cite{HD2}, and \cite{KM} for
discussions of the mathematical ambiguities. It has been observed
\cite{ESCH} that singularities similar to those that are
associated with the Hartle-Hawking transition can also arise in
classical relativity, as in Example 1, above; see also \cite{DR},
\cite{DT}, \cite{DMT}, \cite{H2}, \cite{KK1}, and \cite{KK2}.

\medskip

3. \emph{Repulsive singularities in 4-dimensional extended
supergravity} \cite {GHS}: Mathematically, this model is essentially
a combination of 2\emph{ii)} and 2\emph{iii)};

\medskip

4. \emph{Binary black hole space-times with a helical killing
vector} \cite{Kl}: This model is a generalization of example 1;
see also \cite{HSE} and \cite{To}.

\medskip

5. \emph{Brane worlds}: A \emph{brane} is a submanifold of a
\emph{bulk}, or higher-dimensional space-time. Traditionally,
branes have been represented as uniformly time-like, but it has
been observed \cite{MSV1} that they need not be time-like
everywhere and provide a natural context for signature change. In
fact, mixed Euclidean-Lorentzian branes can be constructed in such
a way that both the bulk and the brane are regular. However, if
viewed from within the brane, the change of signature may appear
as a curvature singularity \cite{GI}, \cite{MSV2}. For certain
choices, this provides an elegant kinematic model for both the
apparent big bang singularity and the apparent accelerated
expansion of the universe; in particular, no hypothesis of dark
energy is needed to account for accelerated expansion \cite{MSV3}.

\medskip

Signature change has also been investigated in the context of spinor
cosmology; see \cite{VJS} for a recent example.

There is a substantial literature on constructing analogies for
the dynamical equations for light in curved space-time using
equations from models of condensed matter, including (but by no
means limited to) Bose-Einstein condensates with a sink or a
vortex \cite{BLSV}, \cite{LVW}, \cite{W1}-\cite{WWVSS}. These lead
to analogies between acoustic waves in matter and wave equations
on mixed Riemannian-Lorentzian manifolds which are reminiscent of
those between gas dynamics and extremal surfaces discussed in
Sec.\ 4.2.

The idea behind these models is elegantly simple: An acoustic wave
has a relation to the flow in which it propagates which is
analogous in some crucial ways to the relation between a light
wave and the ambient space-time. This analogy yields a kinematic
model for certain relativistic effects. For example, if the flow
becomes supersonic, an acoustic wave emitted downstream from a
listener will be trapped in an analogous way to the trapping of
light inside a black hole with respect to an external observer.

Such analogies can be traced back at least to the flow metrics of
\cite{Be} and possibly to the electrodynamics of \cite{Go}; a
review is given in \cite{BLV}.

An elliptic-hyperbolic system associated with the Einstein
evolution equations is studied in \cite{AM}. But in that example
an elliptic gauge-fixing condition is coupled to hyperbolic
evolution equations on a Lorentzian metric. Because the signature
of the metric is fixed, the resulting system is qualitatively
different from the cases considered here.

An important consideration for physical applications is whether
the elliptic-hyperbolic differential operator is of \emph{real
principal type,} in which case the principal symbol of the
operator is real-valued and no complete null bicharacteristic can
be trapped over any compact subset of the domain. Because a null
bicharacteristic is an integral curve of a Hamiltonian system
canonically associated to the principal symbol, the major analytic
properties for operators of real principal type depend only on the
principal symbol, and not on the form of the lower-order terms;
see the concluding remarks in Sec.\ 3 of \cite{Pa2}.

If the operator is of real principal type, ideas from microlocal
analysis can be applied to construct a natural theory of boundary
regularity that is applicable to the elliptic-hyperbolic case
\cite{Pa1}; see also \cite{Gra} and \cite{Gro}. A major difference
between most of the physical examples listed here and the
geometric examples of Secs.\ 1-5 is that the latter operators are
not of real principal type; as a result, microlocal arguments
appear to fail and what can or cannot be said about solutions
tends to depend delicately on the precise form of the lower-order
terms. A typical example is Theorem 3 of \cite{Gu1}. In
particular, this kind of dependence prevents the derivation of
uniqueness theorems by the expected arguments; see Sec.\ 5.1 of
\cite {O12} for an illustration.

\bigskip

\textbf{Acknowledgment}. I am grateful to Prof. K. R. Payne for
acquainting me with the accepted terminology for equations of the
form (\ref{keldysh}). Confusion arises because M. V. Keldysh was
not the first mathematician to study such equations.

\end{document}